\documentclass[journal]{IEEEtran}

\usepackage{hyperref}
\usepackage{amsmath}
\usepackage{amssymb}
\usepackage{multicol}
\usepackage{multirow}
\usepackage{cite}
\usepackage{graphicx}

\usepackage{enumitem}
\ifCLASSINFOpdf
\else

\fi

\begin{document}
%
\title{UnmixingSR: Material-aware Network with Unsupervised Unmixing as Auxiliary Task for Hyperspectral Image Super-resolution}

\author{Yang Yu}

\markboth{Journal of \LaTeX\ Class Files,~Vol.~13, No.~9, September~2014}%
{Shell \MakeLowercase{\textit{et al.}}: Bare Demo of IEEEtran.cls for Journals}

\maketitle

\begin{abstract}
    Deep learning-based (DL-based) hyperspectral image (HIS) super-resolution (SR) methods have achieved remarkable performance and attracted attention in industry and academia. Nonetheless, most current methods explored and learned the mapping relationship between low-resolution (LR) and high-resolution (HR) HSIs, leading to the side effect of increasing unreliability and irrationality in solving the ill-posed SR problem. We find, quite interestingly, LR imaging is similar to the mixed pixel phenomenon. A single photodetector in sensor arrays receives the reflectance signals reflected by a number of classes, resulting in low spatial resolution and mixed pixel problems. Inspired by this observation, this paper proposes a component-aware HSI SR network called UnmixingSR, in which the unsupervised HU as an auxiliary task is used to perceive the material components of HSIs. We regard HU as an auxiliary task and incorporate it into the HSI SR process by exploring the constraints between LR and HR abundances. Instead of only learning the mapping relationship between LR and HR HSIs, we leverage the bond between LR abundances and HR abundances to boost the stability of our method in solving SR problems. Moreover, the proposed unmixing process can be embedded into existing deep SR models as a plug-in-play auxiliary task. Experimental results on hyperspectral experiments show that unmixing process as an auxiliary task incorporated into the SR problem is feasible and rational, achieving outstanding performance. The code is available at
\end{abstract}

\begin{IEEEkeywords}
    Hyperspectral super-resolution, Hyperspectral unmixing, Auxiliary learning.
\end{IEEEkeywords}

%
\IEEEpeerreviewmaketitle

\section{Introduction}

\IEEEPARstart{W}{ith} the rapid development of hyperspectral imaging sensors, hyperspectral images (HSIs) have been broadly applied in many fields \cite{HSIIJCAI,face,abnormal}. Compared with RGB images, HSIs contain more abundant spectral information that delineates objects' components. Nonetheless, mainly due to the limited amount of photons reaching each narrow spectral band, there is a trade-off between the spatial and spectral resolution. HSIs are usually obtained under low spatial resolution and incapable of capturing spatial details, which impedes their practical applications significantly. Therefore, hyperspectral image super-resolution (SR), which is economical and practical without the hardware modification to recover spatial details from the observed HSI, has received wide attention \cite{YQQ}. The existing HSI SR framework can be divided into two categories: fusion-based HSI SR and single HSI SR \cite{IFSR}. Considering the difficulties of obtaining aligned auxiliary images of HSI in reality, we focus on the single HSI SR in this paper \cite{ECCVSR}.

\begin{figure}[h]
    \centering
    \includegraphics[width=\linewidth]{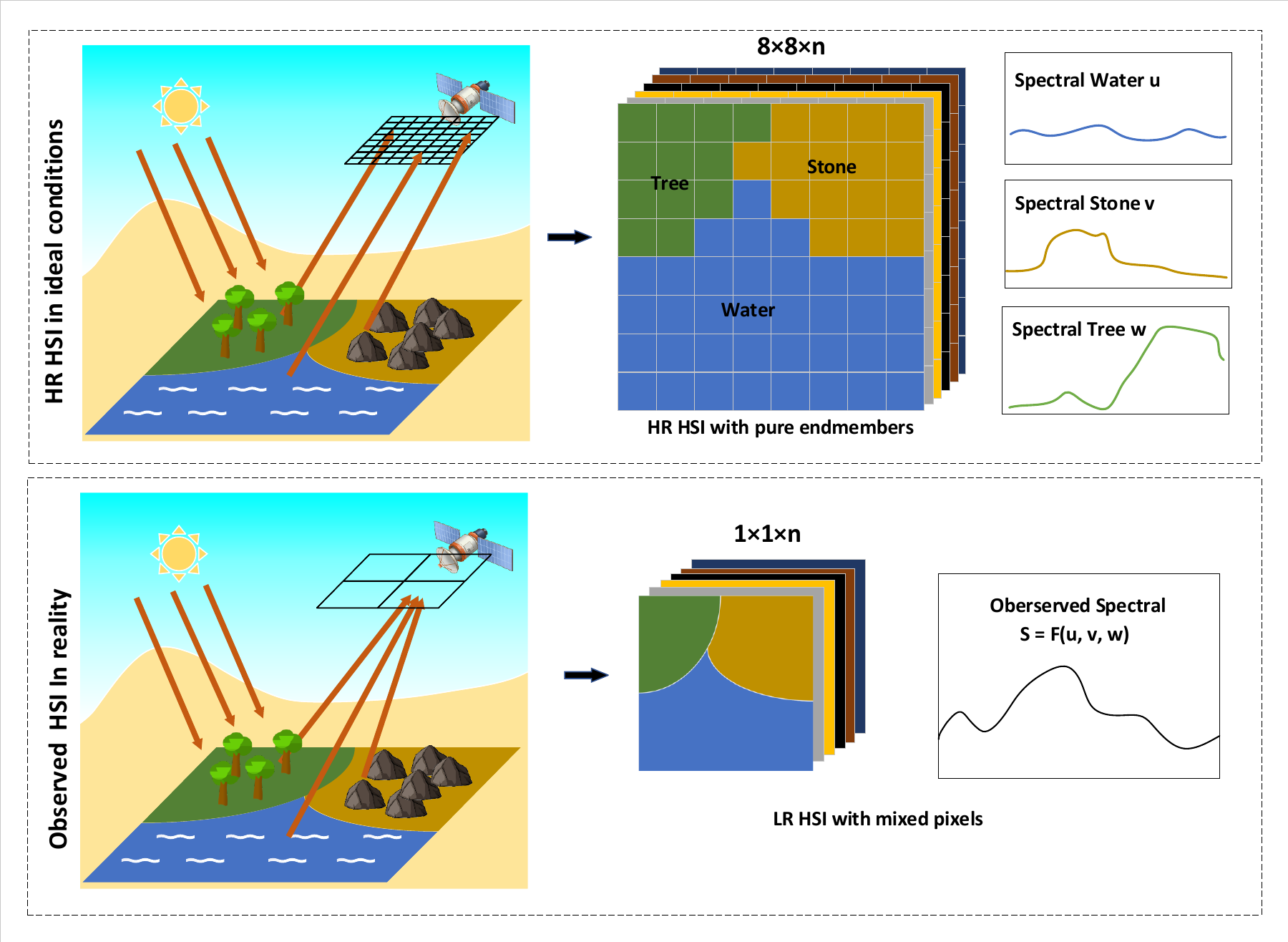}
    \caption{The illustration of the relationship between the LR HSIs and the mixed pixels.}
    \label{mixed}
\end{figure}

Over the past decades, various methods have been proposed for HSI SR problems \cite{deepmatrix,ICCVW,former}. Typically, traditional methods design hand-crafted rules, such as self-similarity, sparsity constraint, or low rank of the HSI, to constrain the SR process. Those rules need sophisticated designing and hard to generalize. Recently, with the development of deep learning (DL), DL-based SR methods have been introduced to solve the HSI SR problem. Those DL-based SR methods learned the relationship between the low-resolution (LR) HSIs and high-resolution (HR) counterparts, achieving preferable performance in solving HSI SR problems. Nonetheless, those DL-based methods, in essence, treat the HSI as a special image with hundreds of spectral bands, ignoring the unique characteristic of HSI different from ordinary RGB images.

From the HSI imaging perspective, we find the relationship between LR HSI and the mixed pixel phenomenon and rethink the HSI SR problem. The issues of the low spatial resolution and mixed pixel problem often coexist, jointly hindering the downstream HSI applications. As shown in Figure \ref{mixed}, in ideal conditions, a photodetector in sensor arrays receives the reflectance signals reflected by a single object. However, in reality, the received signal, mixed with a number of classes, is observed by a photodetector. From the perspective of image quality, this mixed-signal phenomenon leads to the low spatial resolution. From the perspective of spectral characteristics, this phenomenon leads to the spectral profile aliasing, which is also known as the mixed pixel problem. Therefore, the mixed pixel in LR HSI reflects the mixed spatial and spectral features of specific areas. In this situation, considering the mixed problem during the SR process can not only maintain the spectral consistency between the LR-HR images but also effectively improve the spatial performance.

Benefiting from the unique characteristic that HSI reflects subtle spectral properties of objects, the material components and corresponding proportions can be recognized by the unsupervised process in terms of spectral perspective. Hyperspectral unmixing (HU), which decomposes the observed mixed spectrum into the spectral profile of material components (endmembers) and corresponding proportions (abundance maps), is a direct approach to solving the aforementioned problems. The unmixing process reflects the mixing proportion of materials and can be utilized in the SR problem to enhance the overall performance.

Motivated by the relationship between LR HSI and mixed pixels, we regard HU as an auxiliary task and incorporate it into the HSI SR process. Instead of only learning the mapping relationship between LR and HR HSIs, we leverage the bond between LR abundances and HR abundances to boost the stability of our method in solving SR problems. The network starts with the unsupervised unmixing process in LR HSIs.

Introducing additional information shrinks the solution space in solving the ill-posed problem. The material components of HSI can be perceived by decomposing the measured spectrum into its constituent substances (endmembers) and a set of fractional abundances. By perceiving the proportion of material components by HU, it can guide the SR network to produce meaningful results. Moreover,

In summary, the contributions can be summarized as follows:

\begin{enumerate}

    \item Inspired by the observation that LR is similar to the mixed pixel phenomenon in HSIs, we regard HU as an auxiliary task and incorporate it into the HSI SR process by exploring the relationship between LR and HR abundances. The proposed architecture effectively boosts our method's stability in solving SR problems.

    \item Considering the similarity between the SR and unmixing tasks, we design a general residual attention module (GRAM) for both HSI SR and unmixing.

    \item Ablation experiments demonstrate the rationality and feasibility of introducing the unmixing process into the HSI SR problems. As an efficient auxiliary task, the proposed framework can be embedded into existing deep learning architectures to enhance performance effectively.

\end{enumerate}

\section{Related work}

\subsection{DL-based Hyperspectral Image Super-resolution}

Hyperspectral image super-resolution is a fundamental task of hyperspectral processing, aiming at reconstructing high-quality HSIs from low-quality inputs \cite{TIP}. Compared with RGB image SR, HSI SR simultaneously improves spatial resolution and maintains the spectral profile. DL-based methods have naturally introduced and dominated the HSI SR problems for the past few years because of their superior performances in many computer vision tasks \cite{densenet,fcanet}. \cite{MCNet} introduced the mixed 2D/3D convolutional network (MCNet) to extract the potential features. To fully use the HSI spectral information, \cite{wxyunmixing} proposed the dilated projection correction network by transform the HSI SR problem from the image domain into the abundance domain. Considering hyperspectral data's high dimensionality, group convolution is introduced to relieve the learning difficulties. \cite{JJJTCI} proposed the spatial-spectral prior network (SSPSR) by introducing the group convolution (with shared network parameters) and progressive upsampling framework in SR problems. Those DL-based HSI SR methods significantly outperform the traditional methods by learning the mathematical mapping rules of paired HSIs yet neglecting the material constituent relationship between LR and HR HSIs.

\subsection{Auxiliary Learning in HSI SR}

HSI SR is characterized by high-dimensional data, a limited amount of training examples and ill-posed problem. Auxiliary learning, a method to improve the ability of a primary task by training on additional auxiliary tasks alongside this primary task, is a promising framework for solving the challenging HSI SR problem \cite{auxiliarylearning,ECCVML,MTL}. \cite{JJJTIP} proposed a deep spatial-spectral feature interaction network (SSFIN) for spatial-spectral image SR. In SSFIN, the spectral SR task promoted the performance of spatial SR. Considering the absence of HSI SR training data and the plenitude of RGB SR data, some methods introduce the RGB SR as the auxiliary task to solve the HSI SR problem. \cite{WACV} developed a multi-tasking network to train RGB SR and HSI SR jointly so that the auxiliary task RGB SR can provide additional supervision and regulate the HSI SR network training. \cite{ICIP} leveraged an auxiliary RGB learning task to reconstruct the corresponding RGB image from the snapshot image in the training phase and incorporate the learned features of the auxiliary task to assist the more complex reconstruction of the latent HS image. \cite{JSTARSMULTITASK} proposed a multi-scale spatial fusion and regularization-induced auxiliary task to fuse the LR HSI and HR multi-spectral counterpart for reconstructing HR HSI. Auxiliary learning has exhibited great potential in solving the HSI SR problem. However, nearly all methods need extra data to train the auxiliary network. Collecting additional labeled training examples is time-consuming and uneconomical.

\subsection{DL-based Hyperspectral Unmixing}

Hyperspectral unmixing can decompose the high-resolution HSIs into a combination of the material components and corresponding endmembers in an unsupervised manner. Recently, DL-based unmixing methods have attract extensive attentions and achieve superior performance. The DL-based unmixing methods chose the autoencoder as the backbone and add additional reasonable constraints to facilitate the unmixing network. Based on the unmixing autoencoder, researchers have designed effective network architectures to optimize the unmixing performance. For instance, considering the limitation of lacking spatial information in basic autoencoder, CNNAEU incorporate the convolutional neural network into the basic autoencoder. To further capture the multi-scale spatial feature, MSNet poposed the multi-stage convolutional unmixing framework which capture the local-global information by conducting multi-stage sub-networks. Furthermore, transformer choose the transformer as the basic feature extraction block to capture the global information and non-local dependency. Specifically, adversarial learning have been conducted to optimize the unmixing network by capturing the statistical properties of the HSIs. Although these methods have achieved promising performance in traditional unmixing tasks, problems exists when choosing the unmixing as the auxiliary tasks. Traditional unmixing framework taking the single HSI as input and optimize the network training by reconstruction errors. However, in this paper, the unmixing network aims to learn a general network weight for the LR-HR abundance inference. Therefore, the traditional unmixing framework can not fit the auxiliary learning.

\section{Method}

\subsection{Motivation and Formulation}

Auxiliary learning works when a similarity exists between the primary and auxiliary tasks. The improper auxiliary task will produce a negative impact on the overall framework. In this paper, we add unsupervised unmixing as the auxiliary task to enhance the performance and generalization of the HSI SR network. In this section, we discuss the similarity and relationship between unmixing and SR tasks, demonstrating the rationality of our intention.

\begin{description}[style=unboxed,leftmargin=0cm]

    \item[Unmixing task.] For the unmixing problem, we use the linear mixing model as the unmixing framework for its definite physical meaning and convenience of solving. The LMM provides appropriate approximations according to the light scattering mechanisms \cite{JIN}. For the LMM, the HSI $ \mathbf{Y} $  can be linearly combined by the abundance coefficients and endmembers. It can be formulated by Eq. \eqref{eq1}.
        \begin{equation}
            \label{eq1}
            \rm \mathbf{Y = AM}
        \end{equation}
        where  $ \mathbf{YZ} \in \mathbf{R}^{hw \times B}  $ is the LR HSI with $h$ height, $w$ width and $B$ bands. $ \mathbf{A} \in \mathbf{R}^{hw\times p} $ represents the abundance matrix and $ \mathbf{M} \in \mathbf{R}^{p\times B} $  is the endmember matrix with $p$ endmembers of LR HSIs.

    \item[HSI SR task.] The spatial resolution degradation process of HR can be modeled as a convolutional process \cite{PAMIblind,PAMISURVEY}. It can be formulated by Eq. \ref{eq2}.
        \begin{equation}
            \label{eq2}
            \rm \mathbf{Y_{HR} = KY_{LR}}
        \end{equation}
        where $ \mathbf{Y_{HR}} \in \mathbf{R}^{HW \times B} $ is the ground truth (GT) HR HSI with $H$ height, $W$ width and $B$ bands. $ \mathbf{K} $ is the super-resolution kernels. Given the upsampling factor $n$, $H = n \times  h$ and $W = n \times  w$. The HSI SR aims to recover the HR HSI from the LR inputs, i.e., finding the inverse of the degradation kernels.

        \item[Similarity and Relationships]. We can define the HR HSI as a combination of HR abundance maps and endmembers. Therefore, the Eq. \eqref{eq2} can be transformed into Eq. \eqref{eq3}, as follows:
        \begin{equation}
            \label{eq3}
            \rm \mathbf{A_{HR}M_{HR} = K A_{LR}M_{LR}}
        \end{equation}
        where $ \mathbf{A_{LR}} \in \mathbf{R}^{HW\times p}$ is the HR abundances. At the moment of imaging, the substances contained in the images have been fixed. The SR process does not create new substances that does not exist in the corresponding LR images. Therefore, the HR and LR HSI share the same endmembers. Therefore, we can eliminate the endmember items at both ends of the formula and the Eq. \eqref{eq3} can be written as:
        \begin{equation}
            \label{eq4}
            \rm \mathbf{A_{HR} = K A_{LR}}
        \end{equation}
        By eliminating the endmember matrix $\mathbf{M}$ and exploring the relationship between LR abundances and HR abundances, improve the precision of the estimated super-resolution kernel to guide the SR process and boost SR performance.
\end{description}

\begin{figure}[h]
    \centering
    \includegraphics[width=\linewidth]{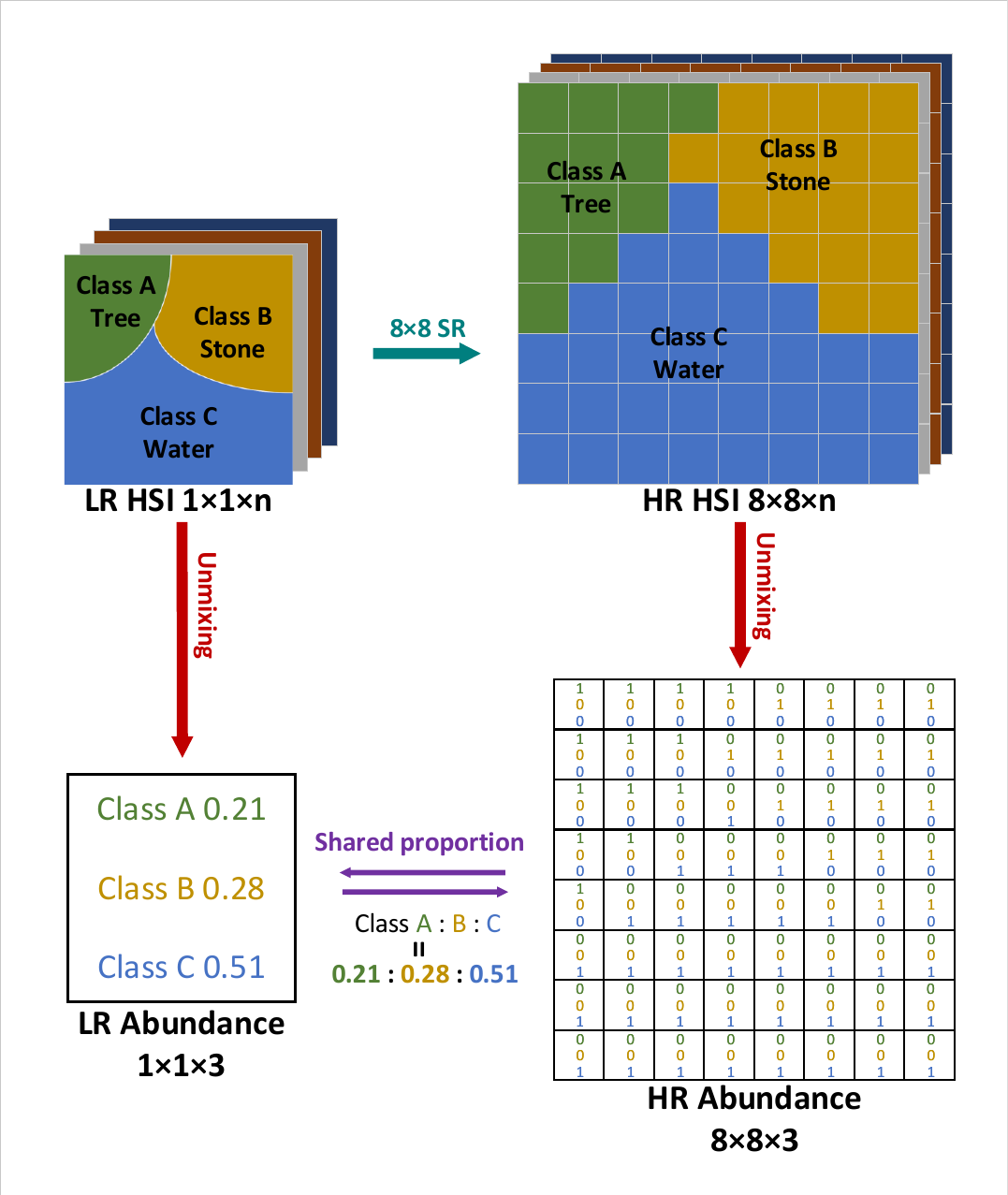}
    \caption{The relationship between the LR-HR abundances.}
    \label{subpixelmapping}
\end{figure}

For the HSI SR task, a single pixel in LR image will be upscaled into $n \times n$ pixels. From the abundance perspective, the counterpart region of LR-HR abundances share the same material proportion. Based on the relationship between the LR-HR images and abundances, we can optimize the SR process. The relationship can be simply represented as Figure. \ref{subpixelmapping}.

\begin{figure}[h]
    \centering
    \includegraphics[width=\linewidth]{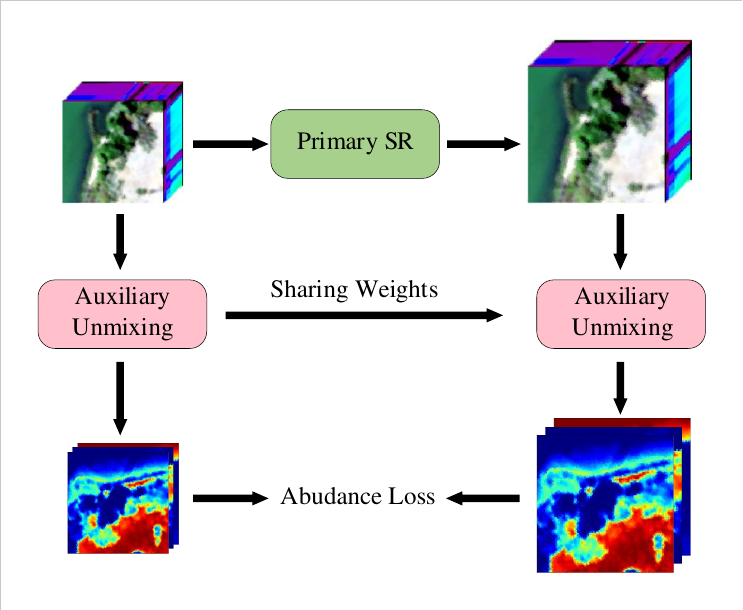}
    \caption{Overview of the UnmixingSR. The proposed framework mainly contains two parts: the primary SR network and the auxiliary HU network. We build the relationship between the primary and auxiliary task by the material-aware module. The auxiliary unmixing network shares the weight. Specifically, the weight of the decoder can be regarded as the endmembers.}
    \label{fig:framework}
\end{figure}

Therefore, the proposed architecture contains a supervised primary HSI SR network and an unsupervised auxiliary unmixing autoencoder network. The primary SR network aims to learn the essential mapping relationship of LR HSI and HR HSI. The auxiliary unmixing network intends to build the constituent relationship between LR abundances and HR abundances, which can help consolidate the HSI SR network. The proportion of material components can be utilized to generate results with significant physical meaning, ensuring the SR network is robust. The overall framework is shown in Figure. \ref{fig:framework}.

\subsection{Auxiliary unsupervised Unmixing autoencoder}
\begin{figure}[h]
    \centering
    \includegraphics[width=\linewidth]{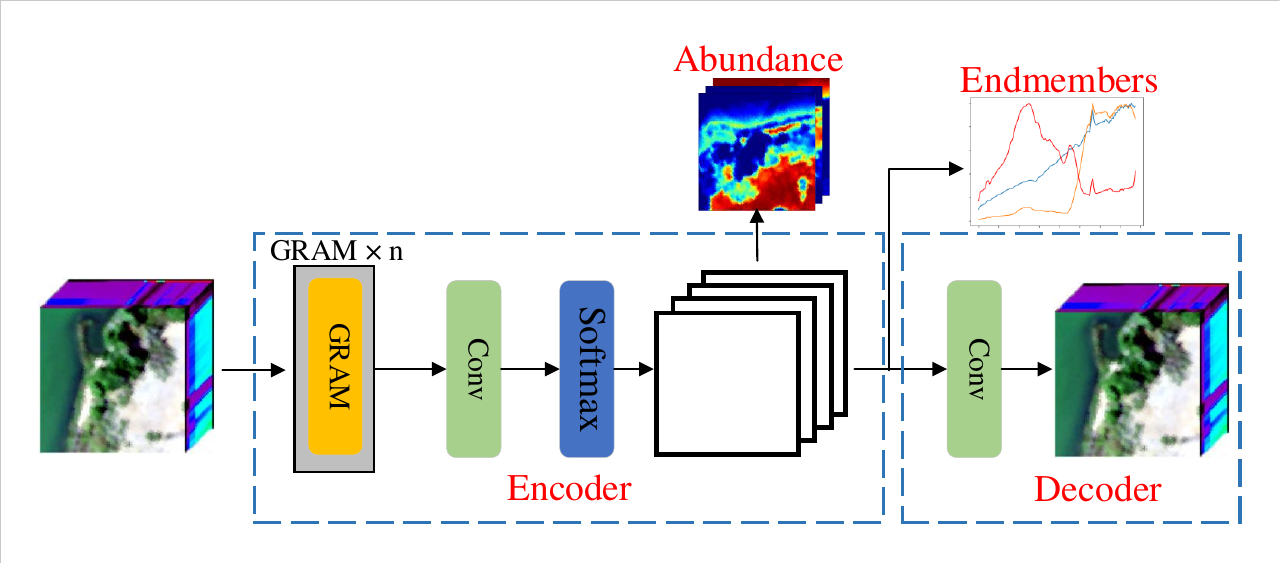}
    \caption{The architecture of the unmixing framework.}
    \label{fig:encoder}
\end{figure}

Regarding the auxiliary unmixing task, we conduct the unmixing autoencoder by stacking several GRAMs for unsupervised hyperspectral unmixing. For the unmixing autoencoder, the abundance maps are obtained from the encoder output whose dimensionality equals the number of endmembers, and the extracted endmembers arise as the decoder layer weights. The endmembers reveal the spectral feature of substances contained in the HSI. The abundance map identifies the proportion of each endmember present in the spectra of each pixel \cite{yuunmixing}. The endmember matrix is subjected to the nonnegative constraint. In addition, the abundance matrix should satisfy the abundance sum-to-one constraint (ASC) and the abundance nonnegative constraint (ANC) at the same time. The unmixing framework is shown in Figure. \ref{fig:encoder}.

The unmixing autoencoder contains two parts: an encoder and a decoder. The encoder contains several GRAMs, convolution layers, and a softmax layer, aiming for feature extraction and dimensional reduction. The encoder ends up with a softmax layer, which aims to ensure the ASC. The abundance maps are obtained from the encoder output, whose dimensionality equals the number of endmembers. The decoder contains a  $1 \times 1$ convolutional layer, which is designed to combine the endmembers and abundances linearly. The $1\times 1$ convolutional decoder layer simulates the LMM, of which the weights represent the extracted endmembers.

For the unsupervised unmixing network, we construct a hybrid loss to achieve desirable unmixing performance. We choose the pixel-wise L1 reconstruction error as the first item to minimize the absolute difference, as shown in \eqref{eql1}. Moreover, to maintain the spectral consistency of the reconstructed HSI, we choose the spectral angle distance (SAD) as the second loss item, as shown in \eqref{eq9}.

\begin{equation}
    \label{eql1}
    \rm UnLoss_{L1} = \parallel \mathbf{Y} - \mathbf{\widehat{Y}} \parallel _{1}
\end{equation}

\begin{equation}
    \label{eq9}
    \rm {UnLoss_{SAD}} = arccos(\frac{\langle \mathbf{Y}, \mathbf{\widehat{Y}}\rangle}{\parallel \mathbf{Y}\parallel _{2} \parallel \mathbf{\widehat{Y}}\parallel _{2}})
\end{equation}
where $\mathbf{Y}$ represents the input HSI, and $\mathbf{\widehat{Y}}$ represents the reconstructed HSI. Specifically, due to the training strategy of the auxiliary unmixing framework, the traditional endmembers initialization methods (such as VCA \cite{vca}) is not applicable in our network. Therefore, we design the spectral total variation loss to preserve the spectral smoothness of the extracted endmembers. It can be formulated by Eq. \eqref{TV}.
\begin{equation}
    \label{TV}
    \rm Loss_{TV}=\sum_i^n(e_{i+1}-e_i)
\end{equation}
where the $\rm e_i$ represents the spectral reflectance of band $\rm i$. In summary, the total loss function of the proposed unmixing is a weighted sum of the three parts:
\begin{equation}
    \label{LossTotal}
    \rm \text{UnLoss}=UnLoss_{L1}+\alpha {UnLoss_{SAD}}+\beta UnLoss_{TV}
\end{equation}
where $\alpha$ and $\beta$ are used to balance the contributions of different terms. We set the $\alpha = 0.1$ and $\beta = 1e-3$ empirically.

\subsection{The framework of the GRAM}

\begin{figure}[h]
    \centering
    \includegraphics[width=\linewidth]{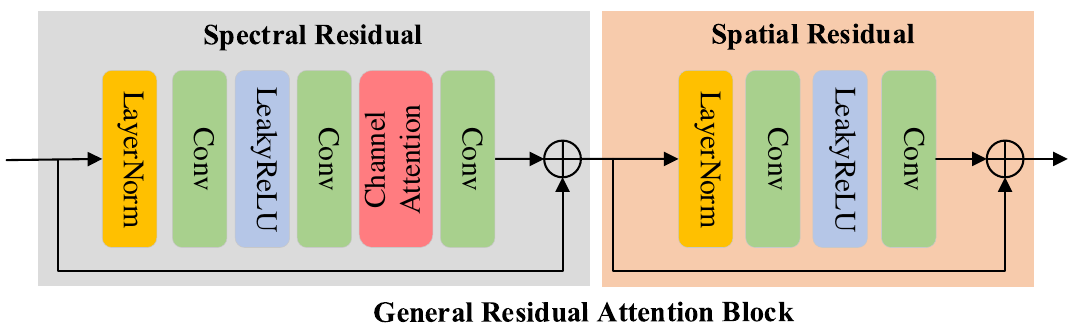}
    \caption{The architecture of the GRAM framework.}
    \label{fig:GRAM}
\end{figure}

Due to the different task objectives and characteristics of super-resolution and unmixing, their feature extraction modules usually have certain differences. In this paper, we designed a general feature extraction framework that can meet the feature extraction requirements of both tasks simultaneously. As shown in Figure \ref{fig:GRAM}, the GRAM consists two parts: the spectral residual block and the spatial residual block. Specifically, the spectral residual block consists of the layer normalization, basic convolutional layer, LeakyReLU activation function and channel attention. The layer normalization process aims to stabilize the feature distribution and accelerate convergence by normalizing the input feature across the channel. Considering the high spectral dimensional and band redundancy of HSI, the channel attention layer aims to capture the channel-wised relationship \cite{cbam}. The spatial residual module consists of a sequence of layer normalization, convolutional layers, ReLU activation functions, and convolutional layers. In addition, the residual learning is introduced to consolidate the transmission of information from input to output.

\subsection{Primary supervised SR network}
\begin{figure}[h]
    \centering
    \includegraphics[width=\linewidth]{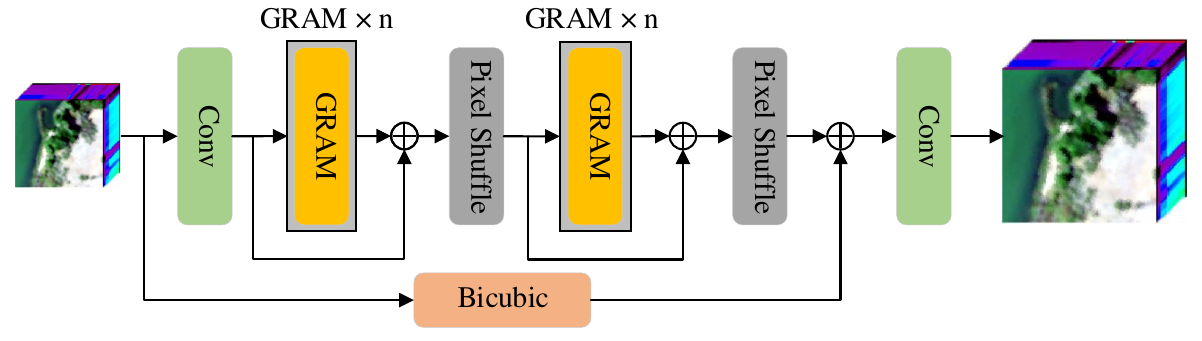}
    \caption{The architecture of the super-resolution framework.}
    \label{fig:SR}
\end{figure}

The architecture of the primary SR network is shown in Fig. \ref{fig:SR}. The network starts with a basic convolution layer to transform the original HSI intro feature domain. After projecting the HSI to the feature domain, we conduct the feature extraction by stacking several GRAMs and achieve feature transfer through residual connections. To decrease the learning difficulty, especially in the large scale SR, we use the progressive upsampling strategy to achieve desirable SR performance. Therefore, the first pixel shuffle module will upscale the feature into $n/2$ ($n$ is the upscaling factor), and the  rest one handles the remaining ×2 factor. Moreover, global residual learning is introduced to consolidate the transmission of information from input to output. Finally, the network work ends with a basic convolution layer and obtains the HR HSI.

For the super-resolution network, similar to the deconvolution network, we construct a hybrid loss consisting of three terms, where the first two terms are L1 loss and SAD loss, as shown in Eq. \eqref{eq7} and \eqref{eq8}, respectively.

\begin{equation}
    \label{eq7}
    \rm SRLoss_{L1} = \parallel \mathbf{Y_{HR}} - \mathbf{{Y_{SR}}} \parallel _{1}
\end{equation}

\begin{equation}
    \label{eq8}
    \rm SRLoss_{SAD} = arccos(\frac{\langle \mathbf{Y_{HR}}, \mathbf{Y_{SR}}\rangle}{\parallel \mathbf{Y_{HR}}\parallel _{2} \parallel \mathbf{Y_{SR}}\parallel _{2}})
\end{equation}
where $\mathbf{Y_{HR}}$ represents the ground truth HR HSI, and $\mathbf{Y_{SR}}$ represents the SR HSI. Moreover, based on the relationship between the unmixing and SR tasks, we construct the abundance loss (AbunLoss) to improve the stability of primary SR task. From the Fig. \ref{subpixelmapping}, it can be observed that the counterpart region corresponding to the LR-HR abundances have the same proportion of material components. Therefore, we constructed a constraint for consistency of component composition in the corresponding regions, as shown in Eq. \eqref{eq11}.

\begin{equation}
    \label{eq11}
    \rm AbunLoss = \parallel \mathbf{A_{SR}} - Deconv(\mathbf{{A_{LR}}}) \parallel _{1}
\end{equation}

The hybrid loss functions of the primary SR network are shown in Eq.\eqref{eq12}.
\begin{equation}
    \label{eq12}
    \rm Loss_{SR} = SRLoss_{L1} + \alpha SRLoss_{SAD} + \beta AbunLoss
\end{equation}
where $\alpha$ and $\beta$ are used to balance the contributions of different terms. We set the $\alpha = 0.1$ and $\beta = 1e-3$ empirically. Compared to the

\subsection{Training Strategy}
The overall framework contains the auxiliary unmixing network and the primary SR network, following the alternate training strategy including two steps. The steps $\rm{I}$ is the LR unmixing training process by minimizing the auxiliary loss function \eqref{LossTotal}. The step $\rm{I}$ aims to perceive the material components and obtains the unmixing network weights. The steps $\rm{II}$ cascades the SR process and the HR unmixing process by minimizing the primary loss \eqref{eq12}. The SR process aims to obtain the HR HSI, and the HR unmixing process aims to obtain the HR abundances for building the LR-HR abundance constraints. The LR-HR abundance constraints introduced in steps $\rm{II}$ can guide SR process and improve the SR stability. It should be noted that the HR unmixing network weights in the steps $\rm{II}$ are shared by the step $\rm{I}$. Additionally, these network weights will remain fixed and are not updated during step $\rm{II}$.

\section{Experiments}
\subsection{Experimental setup}
Regarding the primary SR network, the number of GRAMs is set to $9$ to ensure the receptive fields can cover the entire image and extract deep features. The convolutional kernel size in the SR network is set to $3 \times 3$. The hyper-parameter $\alpha$ is set to $0.1$.

The hyper-parameter of auxiliary loss $\beta$ is set to $0.2$ emphatically.
Our primary and auxiliary tasks share the same learning strategy during the training process. We use the Adam optimizer with $\beta = (0.9,0.99)$ and $weight\_decay = 0$ to optimize the overall network. The learning rate is set to $5 \time 10^{-4}$ and decayed by half every 40 epochs.

\subsubsection{Datasets and comparison methods.}
To evaluate the performance and efficiency of the proposed method, we conduct the experiments on different types of datasets with different upscaling factors. The experimental datasets include the airborne dataset Chikusei and ground hyperspectral dataset ARAD1K, covering various imaging platforms, resolutions, and scenes. For airborne hyperspectral datasets, we chose the Chikusei [70] and Pavia center [71] datasets. The Chikusei dataset contains 2517$\times$2335 pixels with a spatial resolution of 2.5 meters, covering urban and agricultural areas with 128 bands ranging from 363 to 1018 nm. The Pavia Center dataset is acquired by the ROSIS sensor and has a size of 1096 $\times$ 715 pixels, with 103 spectral bands. For the ground hyperspectral dataset, we choose NITRE22 spectral reconstruction dataset (ARAD1K). The ARAD1K dataset comprises 950 ground hyperspectral images, with a size of 512 $\times$ 482 pixels and 31 spectral bands ranging from 400 to 700 nm. The types of ground object in this dataset are very complex and irregular. The dataset contains scenes such as murals, faces, plants, and buildings, with hundreds of material types. We conduct the SR training at $16 \times 16 \rightarrow 64 \times 64 $ for $\times 4$ upscaling and $16 \times 16 \rightarrow 128 \times 128 $ for $\times 8$ upscaling, respectively. We compare our UnmixingSR with five state-of-the-art DL-based HSI SR methods, including MCNet \cite{MCNet}, SSPSR \cite{JJJTCI}, RFSR \cite{RFSR}, aeDPCN \cite{wxyunmixing}, and DeepShare \cite{WACV}.

\subsubsection{Evaluation metrics.}
Four commonly used metrics are used to quantitatively evaluate the HSI SR performance, including Spectral Angle Distance (SAD), Erreur Relative Globale Adimensionnelle de Synthese (ERGAS), Peak signal-to-noise ratio (PSNR), and structural similarity (SSIM). Those metrics reflect the quality of the reconstructed image from different aspects. Specifically, PSNR and ERGAS intend to evaluate the overall quality of the reconstructed HSI. The SSIM evaluates the spatial similarity between GT and estimated HSIs. SAD quantifies the network ability of spectral information preservation, reflecting the similarity between estimated and reference spectral.

\subsubsection{Ablation experiments}

\begin{figure*}[h]
    \centering
    \includegraphics[width=\linewidth]{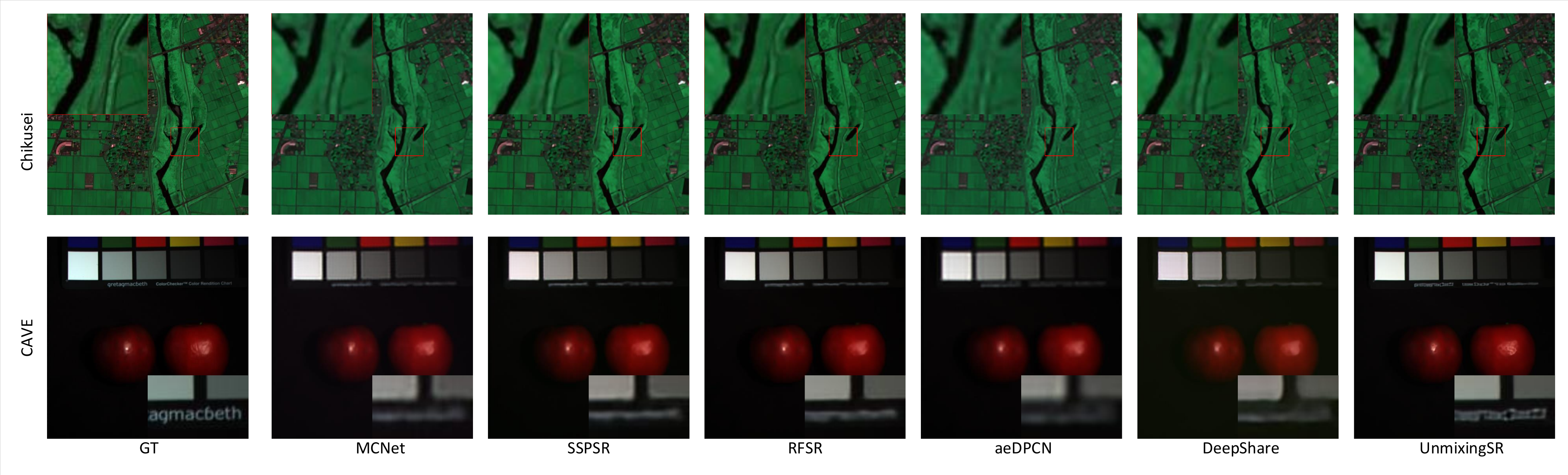}
    \caption{Visualized SR results of methods on Chikusei and CAVE datasets.}
    \label{visualization}
\end{figure*}

\begin{figure*}[h]
    \centering
    \includegraphics[width=\linewidth]{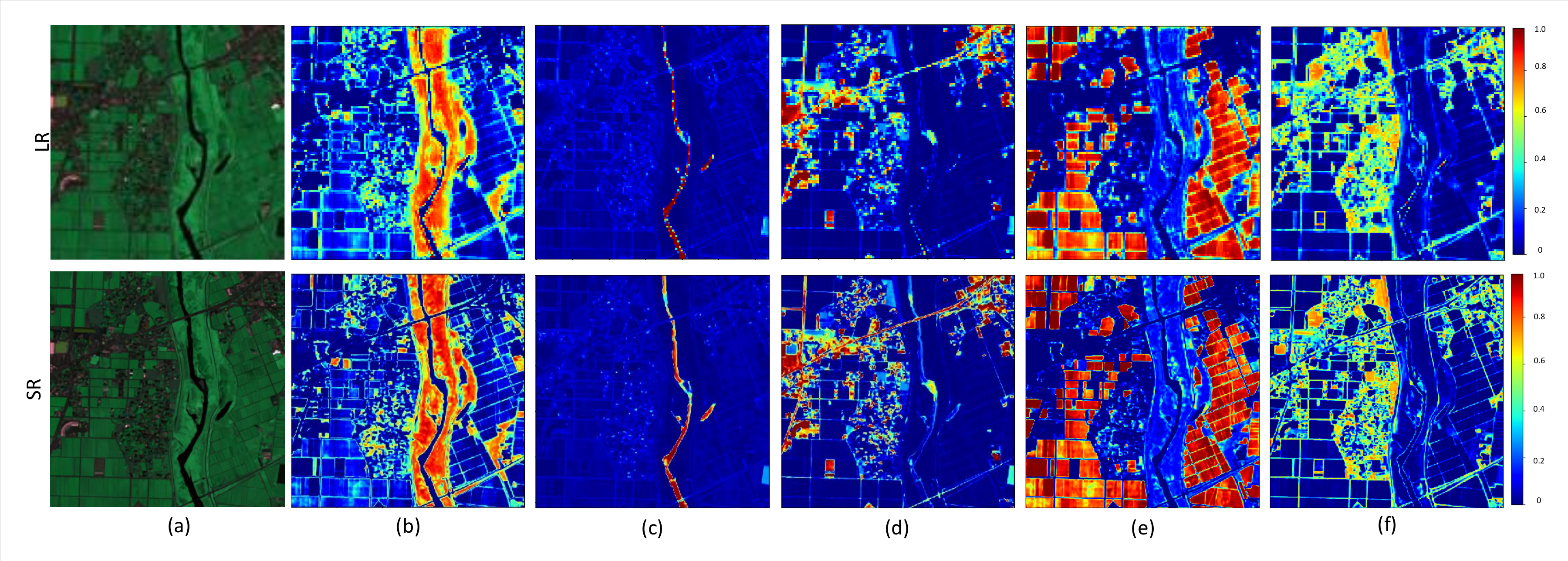}
    \caption{Visualization of abundance maps in auxiliary HU network. (a) represent the LR image and SR HSIs. From (b) to (e), they are the abundance maps obtained by auxiliary HU network.}
    \label{abundance}
\end{figure*}

\begin{figure}[h]
    \centering
    \includegraphics[width=\linewidth]{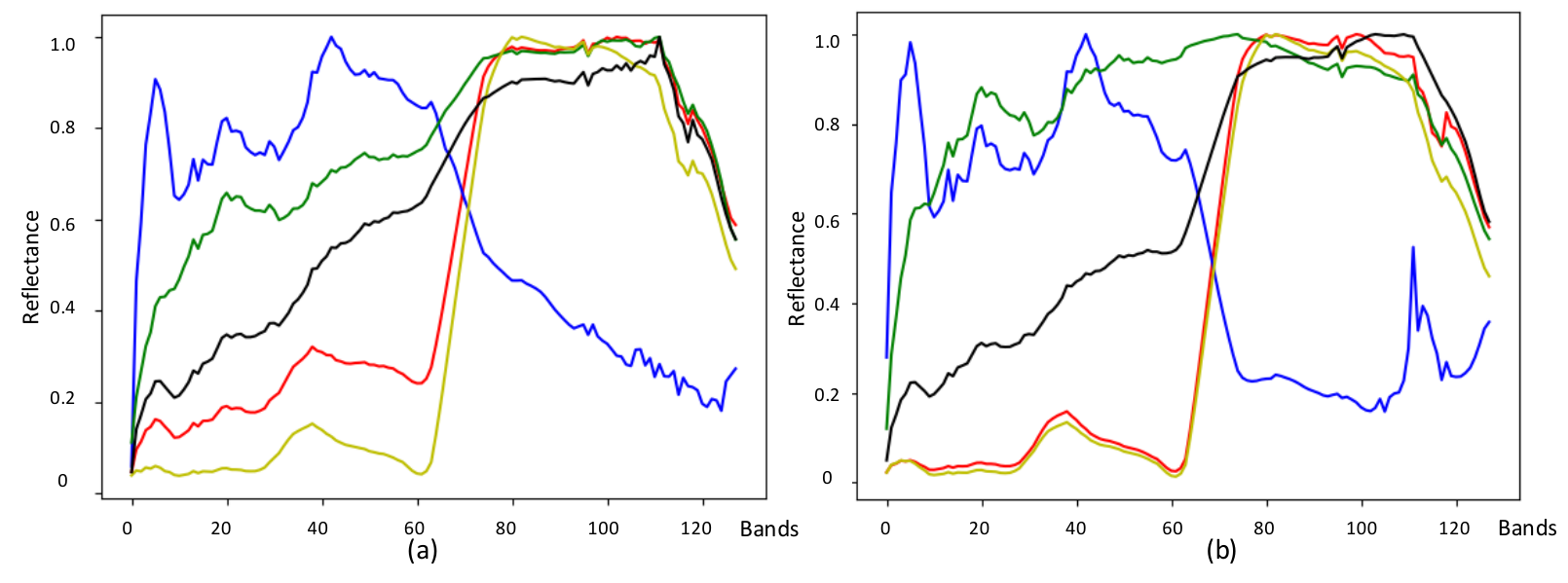}
    \caption{Exctracted endmembers. (a) is the endmembers initialized by VCA and (b) is the extracted endmembers. Different colors represent different substances.}
    \label{endmembers}
\end{figure}

\subsection{Airborne HSI experimental results}

\subsection{Results and Discussions}

\subsubsection{Comparisons of SR results.}

\begin{description}[style=unboxed,leftmargin=0cm]
    \item[Comparisons of SR results.] Table \ref{chiresult} and Table \ref{CAVEresult} show the quantitative results on Chikusei and CAVE datasets, respectively. Our method achieves the best performance in all metrics with different scale factors. The performance improvement of the proposed UnmixingSR mainly contributed to the introduction of unmixing process and the MAM. Our UnmixingSR consolidates the stability of the SR network by building the abundance of interaction between LR and HR HSIs.

        The comparison methods cannot provide promising results because of various reasons and different intentions. With a good result, MCNet can be regarded as a baseline of the CNN HSI SR framework without a sophisticated design. aeDPCN uses unmixing as the preprocessing to transform the SR problem into the abundance domain, which can reduce the computational and spectral distortion. However, the subsequent SR network performs poorly due to the accumulation and propagation of unmixing errors. Moreover, aeDPCN is not an end-to-end network, requiring preprocessing and postprocessing to fine-tune the results. SSPSR introduces the spectral grouping strategy and progressive upsampling scheme to enhance the SR performance. Although the grouping strategy reduces the learning difficulty, it breaks the consistency and correlations of spectral bands. The insufficient exploitation of spectral information limits the reconstruction performance of SSPSR. By modeling the band correlations from a sequence aspect, RFSR achieves competitive performance. DeepShare introduces the RGB SR as an auxiliary task by sharing the spatial SR network to promote the HSI SR performance. To overcome the huge gap in spectral band numbers, DeepShare proposed the divides the HSI into numerous single bands while the RGB into three single bands. Moreover, the sharing network will decrease the HSI SR performance because of the discrepancy in the data distribution between hyperspectral and RGB images. Their grouping strategy also breaks spectral consistency.

        The visualization of SR results is shown in Fig. \ref{visualization}, demonstrating our method's superiority. Benefiting from the reconstructed kernel estimated by the auxiliary HU network, the primary SR network gains noticeable improvement. Consequently, the proposed UnmixingSR reconstruct satisfactory results with more accurate edges and precise structural characteristics, while other methods have unsatisfactory appearances.

    \item[Unmixing Results.] Figure \ref{abundance} and Figure~\ref{endmembers} visualize the unmixing results with abundance maps and endmembers, respectively. We can find valuable information from unmixing results. From LR abundance maps, we can observe the composing proportion of substances corresponding to HR regions. The decoder weights, illuminated in Figure \ref{endmembers}, show the spectral feature of objects. Moreover, the LR-HR abundances exhibit structural similarity with obvious detail differences. The differences between LR-HR abundance represent the degradation process.
\end{description}

\subsection{Unmixing Results}

As we choose the unmixing as the auxiliary task to improve the performance of SR task, the reliability of the unmixing performance will directly affect the SR results. Since there is no unmixing ground truth for the selected datasets, we mainly evaluate the unmixing performance through qualitative assessment.

The results obtained from the ARAD dataset demonstrate that in scenes with highly complex
land cover types, we can effectively represent the image as
a combination of endmembers and abundances, leading to
accurate reconstruction outcomes. These favorable unmixing results suggest that our unmixing network can acquire low-dimensional image representations from the perspective of abundance, thereby exhibiting strong applicability and generalization performance. Thus, the proposed unmixing diffusion model not only suits scenes with a single land cover composition but also performs admirably in scenes featuring complex land cover compositions.

\begin{table}[]
    \caption{Quantitative results of different methods on Chikusei dataset with scale factors $\times$4 and $\times$8.}
    \centering
    \label{chiresult}
    \setlength{\tabcolsep}{2.5mm}{
        \begin{tabular}{cccccc}
            \hline
                       & \multicolumn{1}{c}{Scale} & \multicolumn{1}{c}{SAM$\downarrow $} & \multicolumn{1}{c}{ERGAS$\downarrow $} & \multicolumn{1}{c}{PSNR$\uparrow $} & \multicolumn{1}{c}{SSIM$\uparrow $} \\ \hline
            MCNet      & 4                         & 3.338                                & 6.104                                  & 38.677                              & 0.916                               \\
            SSPSR      & 4                         & 2.513                                & 5.098                                  & 40.104                              & 0.941                               \\
            RFSR       & 4                         & 2.329                                & 4.937                                  & 40.331                              & 0.944                               \\
            aeDPCN     & 4                         & 3.601                                & 6.866                                  & 37.138                              & 0.903                               \\
            DeepShare  & 4                         & 2.908                                & 5.637                                  & 39.293                              & 0.929                               \\ \hline
            UnmixingSR & 4                         & \textbf{2.190}                       & \textbf{4.615 }                        & \textbf{40.622}                     & \textbf{0.951 }                     \\ \hline

            \hline \hline
            MCNet      & 8                         & 4.777                                & 8.650                                  & 36.028                              & 0.845                               \\
            SSPSR      & 8                         & 3.414                                & 7.900                                  & 36.720                              & 0.866                               \\

            RFSR       & 8                         & 3.309                                & 7.693                                  & 36.953                              & 0.873                               \\
            aeDPCN     & 8                         & 4.980                                & 9.108                                  & 35.622                              & 0.832                               \\
            DeepShare  & 8                         & 4.483                                & 8.164                                  & 36.541                              & 0.837                               \\ \hline
            UnmixingSR & 8                         & \textbf{2.919}                       & \textbf{7.412}                         & \textbf{37.311   }                  & \textbf{ 0.886 }                    \\ \hline
        \end{tabular}}
\end{table}

\begin{table}[]
    \caption{Quantitative results of different methods on CAVE dataset with scale factors $\times$4 and $\times$8.}
    \label{CAVEresult}
    \centering
    \setlength{\tabcolsep}{2.5mm}{
        \begin{tabular}{cccccc}
            \hline
                       & \multicolumn{1}{c}{Scale} & \multicolumn{1}{c}{SAM$\downarrow $} & \multicolumn{1}{c}{ERGAS$\downarrow $} & \multicolumn{1}{c}{PSNR$\uparrow $} & \multicolumn{1}{c}{SSIM$\uparrow $} \\ \hline
            MCNet      & 4                         & 3.799                                & 5.009                                  & 38.459                              & 0.967                               \\
            SSPSR      & 4                         & 3.761                                & 4.743                                  & 39.077                              & 0.972                               \\
            RFSR       & 4                         & 3.723                                & 4.799                                  & 39.202                              & 0.974                               \\
            aeDPCN     & 4                         & 4.112                                & 5.411                                  & 37.854                              & 0.960                               \\
            DeepShare  & 4                         & 3.879                                & 5.106                                  & 38.329                              & 0.967                               \\ \hline
            UnmixingSR & 4                         & \textbf{3.533 }                      & \textbf{4.437}                         & \textbf{39.891}                     & \textbf{0.979}                      \\ \hline

            \hline \hline
            MCNet      & 8                         & 5.809                                & 9.274                                  & 31.967                              & 0.893                               \\
            SSPSR      & 8                         & 4.665                                & 8.397                                  & 33.610                              & 0.916                               \\
            RFSR       & 8                         & 4.484                                & 8.203                                  & 33.799                              & 0.919                               \\
            aeDPCN     & 8                         & 6.215                                & 9.707                                  & 31.073                              & 0.871                               \\
            DeepShare  & 8                         & 5.408                                & 8.761                                  & 33.158                              & 0.913                               \\ \hline
            UnmixingSR & 8                         & \textbf{4.356 }                      & \textbf{7.911 }                        & \textbf{34.431}                     & \textbf{0.927 }                     \\ \hline
        \end{tabular}}
\end{table}

\subsection{ Ablation Study}

In this section, ablation experiments are conducted to demonstrate the feasibility and effectiveness of the auxiliary unmixing process and MAM in the HSI SR network. Specifically, we integrate the unmixing process into the existing Deep SR approaches to investigate the generalizability of introducing unmixing as an auxiliary task. We conduct the ablation experiments on the Chikusei dataset with scale factor 4. Table \ref{ablation} illustrates the quantitative comparisons between methods with and without unmixing process.

We take UnmixingSR without material-aware module (MAM) and unmixing process as the basic framework to investigate the effectiveness of the proposed module. The proposed framework productively boosts the performance in spectral and spatial metrics. The PSNR, SSIM, and SAM are improved by 0.21, 0.07, and 0.005, respectively. The comparison results demonstrate the effectiveness of the proposed method.

In generalizability experiments, considering the special framework of aeDPCN (unmixing as a preprocessing) and DeepShare (RGB SR as an auxiliary task), we only investigate the impacts of the unmixing process for other methods. Because of the differences in upsampling positions and methods between networks, the MAM module only plugs into the last upsampling layer. The network embedded with the unmixing task performs better than the single one. The comparison results demonstrate the generalizability of plugging the unmixing task into the HSI SR network.

\begin{table}[]
    \caption{The impact of auxiliary unmixing process in HSI SR problem.}
    \centering
    \label{ablation}
    \setlength{\tabcolsep}{2.5mm}{

        \begin{tabular}{ccccc}
            \hline
                                                            & \multicolumn{1}{c}{Unmixing} & \multicolumn{1}{c}{PSNR} & \multicolumn{1}{c}{SAM} & \multicolumn{1}{c}{SSIM} \\ \hline
            \multirow{2}{*}{MCNet}                          & $\times$                     & 38.677                   & 3.338                   & 0.916                    \\
                                                            & \checkmark                   & 38.814                   & 3.281                   & 0.918                    \\ \hline
            \multirow{2}{*}{SSPSR}                          & $\times$                     & 40.104                   & 2.513                   & 0.941                    \\
                                                            & \checkmark                   & 40.157                   & 2.500                   & 0.942                    \\ \hline
            \multicolumn{1}{l}{\multirow{2}{*}{RFSR}}       & $\times$                     & 40.331                   & 2.329                   & 0.944                    \\
            \multicolumn{1}{l}{}                            & \checkmark                   & 40.384                   & 2.322                   & 0.944                    \\ \hline
            \multicolumn{1}{l}{\multirow{2}{*}{UnmixingSR}} & $\times$                     & 40.414                   & 2.261                   & 0.946                    \\
            \multicolumn{1}{l}{}                            & \checkmark                   & 40.622                   & 2.190                   & 0.951                    \\ \hline
        \end{tabular}}
\end{table}

\section{ Conclusion}
In this paper, we propose a novel HSI SR framework that incorporates unsupervised hyperspectral unmixing as the auxiliary task. Inspired by the similarity between LR imaging and mixed pixel phenomenon, we find the relationship between LR and HR abundances which are leveraged to consolidate the stability of solving the ill-posed HSI SR problem. The extra information captured by abundances is introduced into the SR network via the proposed material-aware module (MAM), guiding the network to produce SR results with definite physical significance. Benefiting from the rational design of the primary SR task and auxiliary HU task, our method achieves state-of-the-art results. Moreover, experiments demonstrate that the proposed unmixing process can be integrated into existing deep HSI SR models as an efficient auxiliary task.

\newpage
\newpage

\bibliographystyle{ieeetr}
\bibliography{ref.bib}

\ifCLASSOPTIONcaptionsoff
    \newpage
\fi
\end{document}